\documentclass[12pt]{article}
\usepackage{epsf}
\def\be{\begin{equation}}
\def\ee{\end{equation}}
\def\bea{\begin{eqnarray}}
\def\eea{\end{eqnarray}}

\begin{document}
\begin{titlepage}
\begin{center}
{\Large \bf William I. Fine Theoretical Physics Institute \\
University of Minnesota \\}
\end{center}
\vspace{0.2in}
\begin{flushright}
FTPI-MINN-16/01 \\
UMN-TH-3512/16 \\
January 2016 \\
\end{flushright}
\vspace{0.3in}
\begin{center}
{\Large \bf Light Quark Spin Symmetry in  $Z_b$ Resonances? 
\\}
\vspace{0.2in}
{\bf  M.B. Voloshin  \\ }
William I. Fine Theoretical Physics Institute, University of
Minnesota,\\ Minneapolis, MN 55455, USA \\
School of Physics and Astronomy, University of Minnesota, Minneapolis, MN 55455, USA \\ and \\
Institute of Theoretical and Experimental Physics, Moscow, 117218, Russia
\\[0.2in]

\end{center}

\vspace{0.2in}

\begin{abstract}
It is argued that the recent Belle data, consistent with no activity in the spectrum of the $B^* \bar B + B \bar B^*$ pairs at the mass of the $Z_b(10650)$ resonance, imply that the part of the interaction between heavy mesons that depends on the total spin of the light quark and antiquark is strongly suppressed. In particular this part appears to be significantly weaker than can be inferred from pion exchange. If confirmed by future more detailed data, the symmetry with respect to the light quark spins, in combination with the heavy quark spin symmetry, would imply existence of four additional $I^G=1^-$ resonances at the thresholds for heavy meson-antimeson pairs.

\end{abstract}
\end{titlepage}

Numerous experimental studies at the onset of heavy flavor thresholds have uncovered a rich variety of quarkonium-like states with  many of them presenting a challenge for understanding their internal dynamic structure. In particular, some of the observed peaks very near thresholds for heavy meson-antimeson pairs are likely resonant or bound `molecular'~\cite{ov} states of such pairs that arise largely due to the interaction between the light components of the heavy mesons. For this reason further studies of the near threshold behavior of the heavy meson pairs can provide an insight into new details of the hadronic strong interaction. 

A very clean example of threshold states is presented by the $Z_b(10610)$ and $Z_b(10650)$ peaks found~\cite{bellez} by the Belle experiment in the decays $\Upsilon(5S) \to Z_b \, \pi$. These states should undoubtedly contain a light quark-antiquark pair in addition to the heavy quark $b \bar b$ pair, since they come in full isotopic triplets containing the electrically charged ones $Z_b^\pm$ and the neutral $Z_b^0$~\cite{bellez0}. The masses of the $Z_b(10610)$ and $Z_b(10650)$ resonances coincide within few MeV with the thresholds for pairs of heavy mesons: $B^* \bar B$ and $B^* \bar B^*$ respectively. In fact, at the present accuracy it is not known whether the resonances are below, at, or above the corresponding thresholds. The observed properties of the resonances agree well with the model~\cite{bgmmv} describing them as molecular states made of the corresponding heavy meson-antimeson pair in the $I^G(J^P)=1^+(1^+)$ state~\footnote{It should be mentioned that alternative models for the $Z_b$ resonances have been suggested, in particular a tetraquark model~\cite{ampr,ahw} based on a scheme~\cite{mppr} with diquark correlations.}. In particular, this model explains the apparent strong violation of the Heavy Quark Spin Symmetry (HQSS) in the $Z_b$ resonances, which manifests itself in a comparable rate of their decay into ortho-bottomonium states, $Z_b \to \Upsilon(nS) \, \pi$ ($n=1,2,3$) and into para-bottomonium $Z_b \to h_b(kP) \, \pi$ ($k=1,2$). Indeed, in each of the heavy mesons in a widely separated meson pair the spin of the $b$ quark ($\bar b$ antiquark) is correlated with the spin of the light antiquark (light quark), so that the system is in a mixed state with respect to the total spin of the $b \bar b$ pair. Namely, the spin structure of the meson-antimeson pairs in terms of the total spin of the $b \bar b$ and $q \bar q$ pairs is~\cite{bgmmv}
\bea
&&Z_b(10610) \, \sim  \,\left | B^* \bar B, B \bar B^* \right \rangle_{I^G(J^P)=1^+(1^+)} \, \sim \, {1 \over \sqrt{2}} \left ( 1^-_{b \bar b}\otimes 0^-_{q \bar q} + 0^-_{b \bar b}\otimes 1^-_{q \bar q} \right )~, \nonumber \\
&&Z_b(10650) \, \sim \, \left | B^* \bar B^*\right \rangle_{I^G(J^P)=1^+(1^+)} \, \sim \, {1 \over \sqrt{2}} \left ( 1^-_{b \bar b}\otimes 0^-_{q \bar q} - 0^-_{b \bar b}\otimes 1^-_{q \bar q} \right )~.
\label{zspin}
\eea
This spin structure of the $Z_b$ resonances  predicts a similarity between the resonances, namely, the relations $M[Z_b(10650)]-M[Z_b(10610)] \approx M(B^*) - M(B)$, $\Gamma[Z_b(10650)] \approx \Gamma[Z_b(10610)]$ and also certain amplitude and phase relations between the amplitudes of the decays involving the $Z_b$ resonances~\cite{bgmmv}. The presently available data generally agree, within the existing uncertainties,  with these predictions, although for the amplitude relations between the decays of the $Z_b$ states to bottomonium and a pion the agreement is only within  two sigma (a recent discussion can be found e.g. in Ref.~\cite{hkmmnw}).

It should be noted, however,  that the heavy-light spin structure, described by Eq.(\ref{zspin}) is inferred from the that for free pairs of heavy mesons. In order for this description to be fully applicable to the interacting meson pairs, especially when the interaction results in a threshold singularity, it also has to be assumed  that, in addition to HQSS, the interaction between the mesons does not depend on the total spin of the light quark-antiquark pair. Indeed, if the interaction in the state $1^-_{q \bar q}$ was different from that in the state $0^-_{q \bar q}$, the two states in Eq.(\ref{zspin}) would not be the eigenstates for interacting mesons. The fact that the predictions from Eq.(\ref{zspin}) are quite close to the experimental observations is thus nontrivial and remarkable and possibly indicates that there is an approximate Light Quark Spin Symmetry (LQSS) in the interaction. 

Furthermore, the decay $Z_b(10650) \to B^* \bar B + B \bar B^*$ of the heavier resonance  into the lighter pairs of mesons, perfectly allowed by any conservation laws,  is forbidden in the limit of exact LQSS, and can thus be used as a measure of the dependence of the forces between heavy mesons on the spin of light quarks. Recently Belle has reported~\cite{bellebb2} an updated analysis of the spectra of the $B^* \bar B + B \bar B^*$ and $B^* \bar B^*$ pairs in the decays $\Upsilon(5S) \to B^{(*)} \bar B^{(*)} \, \pi$. In particular, besides the expected [from Eq.(\ref{zspin})] decays $Z_b(10610) \to B^* \bar B + B \bar B^*$ and $Z_b(10650) \to B^* \bar B^*$, their results show no features (above the uncertainties) in the spectrum of invariant mass in the channel  $B^* \bar B + B \bar B^*$ at the mass of $Z_b(10650)$ that would indicate a presence of a coupling of this channel to the latter higher resonance. This (non)observation indicates that LQSS does indeed work for heavy mesons better than expected. 

An immediate consequence of the similarity of interaction in the $0^-_{q \bar q}$ and $1^-_{q \bar q}$ states of the light quarks in a pair of heavy mesons would be an existence of additional molecular resonances at all three thresholds for the the $B^{(*)} \bar B^{(*)}$ pairs. Indeed, by combining the  spin states of the light quarks with $0^-_{b \bar b}$ and $1^-_{b \bar b}$ one finds~\cite{mv11}, in addition to the two $Z_b$ states with $I^G(J^P)=1^+(1^+)$, also four isovector states with negative $G$ parity: two states with $J^P=0^+$ and $2^+$ at the $B^* \bar B^*$ threshold, one  $J^P=1^+$ state at the $B^* \bar B$ threshold, and one $0^+$ state at the threshold for $B \bar B$. The interaction between the mesons does not depend on the spin state of the heavy quark pair due to HQSS. If this interaction also does not depend on the light quark spins, the binding in all six channels should be the same. Due to their negative $G$ parity the additional resonances cannot be produced in a single pion emission in $e^+e^-$ annihilation, however they can be observed in radiative transitions and in processes with emission of $\rho$ in $e^+e^-$ annihilation at higher energy~\cite{mv11}.

It is quite clear that the status of the HQSS and LQSS is completely different. Namely, the HQSS follows from the underlying QCD and the parameter for this symmetry for the $b$ quarks is $\Lambda_{QCD}/m_b$. The LQSS apriori has no theoretical justification.
In particular, LQSS is manifestly broken by the pion exchange between the heavy mesons. The purpose of this paper is to consider the behavior of the $B^* \bar B + B \bar B^*$ invariant mass spectrum near the $Z_b(10650)$ resonance in the presence of an LQSS breaking potential $V$. This behavior is sensitive to the potential $V(q)$ in momentum space at the momentum $q \approx \sqrt{M(B) \, \Delta} \approx 0.5 \,$GeV with $\Delta = M(B^*) - M(B) \approx 45\,$MeV. 
The experimental uncertainty in the existing data is still large for a placing a stringent bound on the spin-dependent potential. However, it can already be argued that a full-strength pion exchange would result in an unacceptably large effect, and that a certain suppression is required in order to interpret the data of Ref.~\cite{bellebb2}. 

In what follows it is assumed that the breaking of LQSS by a spin-dependent potential $V$ can be treated as a perturbation over the LQSS limit. It will also be assumed, as indicated by the data, that all the heavy meson pairs in the decay $\Upsilon(5S) \to B^{(*)} \bar B^{(*)} \, \pi$ emerge from  the $Z_b$ resonances with no non-resonant background. The data~\cite{bellebb2} are quite consistent with these assumptions. Indeed, the pion in the decay $\Upsilon(5S) \to Z_b \, \pi$ is emitted in the $S$ wave, and the chiral symmetry requires that the amplitude is proportional to the pion energy. Moreover, in the HQSS and LQSS limit the coupling in this amplitude is the same for both $Z_b$ resonances. Thus one can estimate the relative yield of the two types of the heavy meson pairs as
\be
{\Gamma[\Upsilon(5S) \to Z_b(10610) \, \pi \to (B^* \bar B + B \bar B^*) \pi] \over \Gamma[ \Upsilon(5S) \to Z_b(10650) \, \pi \to B^* \bar B^*  \pi]} \approx {p_1 \, E_1^2 \over p_2 \, E_2^2} {{\cal B}[Z_b(10610) \to B^* \bar B + B \bar B^*] \over {\cal B}[Z_b(10610) \to B^* \bar B*]} \approx 2.3~,
\label{yr}
\ee
where $p_1$ and $p_2$ ($E_1$ and $E_2$) are the values of the pion momentum (energy) in the corresponding transition, and the branching fractions for the $Z_b$ decays are the measured ones~\cite{bellebb2}: ${\cal B}[Z_b(10610) \to B^* \bar B + B \bar B^*] = (82.6 \pm 2.9 \pm 2.3) \% $ and  ${\cal B}[Z_b(10610) \to B^* \bar B*] = 70.6 \pm 4.9 \pm 4.4 \%$. The estimate (\ref{yr}) agrees\footnote{Thus the result of the improved analysis~\cite{bellebb2} of the data does not support an earlier indication~\cite{hkmmnw} of a relative suppression for the coupling in $\Upsilon(5S) \to Z_b(10650) \, \pi$ in comparison with $\Upsilon(5S) \to Z_b(10610) \, \pi$.} with the directly measured ratio of the yield of the heavy meson pairs: $2.0 \pm 0.4$. (The relation (\ref{yr}) assumes the limit of small width of both resonances. Allowing for the finite width and the shape of the excitation in the heavy meson-antimeson channel would introduce additional model dependence which appears to be premature, given the current experimental uncertainty.)

Then in the first order in the LQSS breaking potential $V$ the production of the channel $B^* \bar B + B \bar B^*$ proceeds through the processes shown in Fig.~1. The effect of the channel mixing in Fig.~1b is the largest at the peak of the $Z_b(10650)$ resonance which practically coincides with the threshold for $B^* \bar B^*$. Thus the contribution of the mechanism of Fig.~1b can be considered at the c.m. energy $E$ of the heavy meson pair close to this threshold at $2 \, M(B^*)$ and the c.m. momentum $k$ of the $B^*$ mesons, $E- 2 M(B^*) = k^2/M(B^*)$, can be treated in the lowest order. In such treatment the momentum transfer $q$ in the rescattering process can be fixed to its threshold value $q \approx 0.5\,$GeV. The energy transfer at this point is small, $q_0 = \Delta/2$, and can be neglected. 

\begin{figure}[ht]
\begin{center}
 \leavevmode
    \epsfxsize=16cm
    \epsfbox{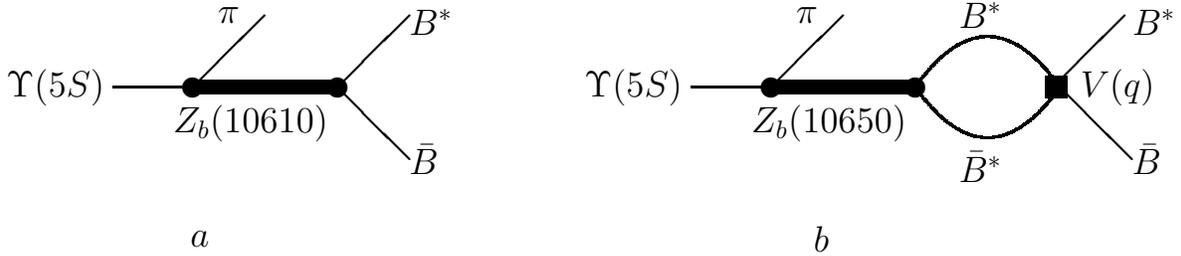}
    \caption{Production of the $B^* \bar B$ pairs through the $Z_b(10610)$ resonance $(a)$ and through rescattering induced by the spin-dependent potential $V$ from the $B^* \bar B^*$ pairs produced through $Z_b(10650)$ $(b)$.  }
\end{center}
\end{figure}

Only the absorptive part of the meson loop in Fig.~1b can be found reliably from the unitarity on-shell cut, while the dispersive part, determined by off-shell contribution, is not calculable at present. The latter part however is a smooth function of $k^2$ and can be approximated by a constant. After a straightforward calculation one can find the amplitude for the production of the $B^* \bar B + B \bar B^*$ channel in the form
\be
A(B^* \bar B + B \bar B^*) = A_1(E) - A_2(E) \, (\mu + i \, k) \, {M \, V(q) \over 2 \, \pi}~,
\label{abb}
\ee
where $M=M(B^*)$, $\mu$ is a constant (of order $\Lambda_{QCD}$) parametrizing the dispersive part of the loop in Fig.~1b, and $A_1$ and $A_2$ stand for the resonant amplitudes
\be
A_1 = A[\Upsilon(5S) \to Z_b(10610) \, \pi \to (B^* \bar B + B \bar B^*) \pi]~, ~~~A_2= A[ \Upsilon(5S) \to Z_b(10650) \, \pi \to B^* \bar B^*  \pi]~.
\label{a12}
\ee
Finally, the part of the amplitude (\ref{abb}) proportional to $k$ is the absorptive part unambiguously determined from unitarity. Using the expression (\ref{abb}), the rate of production of the $B^* \bar B + B \bar B^*$ pair can be written in terms of the pure resonant rate through the $Z_b(10610)$ and a multiplicative factor $R$:
\be
{d \Gamma[\Upsilon(5S)  \to (B^* \bar B + B \bar B^*) \pi] \over dE}= { d \Gamma[\Upsilon(5S) \to Z_b(10610) \, \pi \to (B^* \bar B + B \bar B^*) \pi] \over dE} \, R(E)~,
\label{rdef}
\ee
with $R(E)$ given by
\be
R= \left | 1- {A_2(E) \over A_1(E)} \, (\mu + i \, k) \, {M \, V(q) \over 2 \, \pi} \right |^2~.
\label{rexp}
\ee
The amplitudes $A_1$ and $A_2$ can be taken in the Breit-Wigner form with the coefficients being the same in the symmetry limit. [And, as discussed above in connection with Eq.(\ref{yr}), this limit is in a reasonable agreement with the existing data.] Thus the ratio of the resonant amplitudes in Eq.(\ref{rexp}) takes the form
\be
{A_2(E) \over A_1(E)} = e^{i (\delta_2 - \delta_1)} \, {E - M_1 + i \, \Gamma_1/2 \over E - M_2 + i \, \Gamma_2/2}~,
\label{ramp}
\ee
with the indices 1 and 2 denoting respectively the parameters of the $Z_b(1061)$ and $Z_b(10650)$ resonances, and $\delta_1$ and $\delta_2$ being the non-resonant scattering phases for the corresponding channels.

The non-resonant scattering phases are slowly varying in the energy range around $E = M_2$ considered here, and their difference in Eq.(\ref{ramp}) can be approximated by a constant within the width $\Gamma_2$ at the $Z_b(10650)$ resonance. As a result the effect of the rescattering on the production of the $B^* \bar B + B \bar B^*$ pairs near this resonance is described by three real constants: $\delta_2 - \delta_1$, $\mu$ and $V(q)$, all of which can in principle be determined from the data, provided that a more accurate experimental information becomes available. 

For an approximate estimate, based on the existing data~\cite{bellebb2}, one can notice that the difference of the non-resonant phases is not expected to be large, and can be neglected altogether. The main effect in the factor $R(E)$ is the rapidly varying proportionally to $k$  absorptive part of the rescattering amplitude. This effect is maximal and linear in $V$ when $E$ is within $\Gamma_2/2 \approx 6\,$MeV from the $Z_b(10650)$ resonance, i.e. at $k$ less than about 180\,MeV, and the parameter for the deviation of $R$ from one is approximately given by
\be
R-1 \approx - {2 \, \Delta \over \Gamma_2} {M \, k \, V(q) \over \pi} \approx - {k \over 180\,{\rm MeV}} \, {V(q) \times 2.3 \, {\rm GeV}^2}~.
\label{dr}
\ee

Within the uncertainties, the available data possibly allow for a variation of the yield of $B^* \bar B + B \bar B^*$ around the invariant mass 10.65\,GeV, comparable with the yield itself (already small in comparison with the maximum at low invariant mass due to the $Z_b(10610)$ resonance). However no major feature is present in the data on the spectrum at that energy. Thus it is likely  appropriate to conclude that $q \approx 0.5\,$GeV the spin-dependent potential $V(q)$ should not be much larger than the inverse of $2.3 \, {\rm GeV}^2$.  

It is instructive to compare the latter estimate with the only a somewhat reliably known spin-dependent interaction between the heavy mesons, namely the one generated by pion exchange. The interaction of the mesons with the pion triplet $\pi^a$ is described by the effective Hamiltonian
\be
H_{int} = {g \over f_\pi} \, \left \{ \left [ (B^*)^\dagger_l \tau^a B + {\rm h.c.} \right ]+ i \, \epsilon_{ljk} \, (B^*)^\dagger_j \tau^a B^*_k \right \}  \,  \partial_l \pi^a~,
\label{hint}
\ee
where a nonrelativistic normalization is implied for the heavy mesons, and $\tau^a$ are the Pauli matrices. The charged pion decay constant $f_\pi \approx 132\,$MeV is used for normalization, and the dimensionless constant can be evaluated using the heavy quark limit and the known~\cite{babar,pdg} rate of the decay $D^{*+} \to D \pi$: $g^2 \approx 0.15$. The pion exchange between the mesons in the $I^G(J^P)=1^+(1^+)$ channel then results in the mixing potential~\cite{nv}
\be
V(q)= { 2 \, g^2 \over 3 \, f_\pi^2} \, {q^2 \over q^2+ m_\pi^2}~.
\label{vpi}
\ee 
This potential, if used without a damping form factor,  corresponds to the factor in Eq.(\ref{dr}) $V(q) \times 2.3 \, {\rm GeV}^2 \approx 12$, and, even with the uncertainty inherent in the present estimates of the effect, would give rise to a very significant variation of the yield in the $B^* \bar B + B \bar B^*$ channel at the invariant mass in the vicinity of the higher resonance $Z_b(10650)$ that likely would not be compatible with the data~\cite{bellebb2}. It can be noted that interaction arising from exchange of other pseudoscalar mesons, the $\eta$ (recently considered in detail in Ref.~\cite{kr}) and $\eta'$, does not cancel by itself the effect of the pion exchange at $q \approx 0.5\,$GeV due to larger masses of those mesons. Thus a suppression of the force induced by the pion exchange has to come from a form factor in the interaction (\ref{vpi}). 

Although the apparent approximate LQSS is quite unexpected, if established for the heavy meson-antimeson pairs with quantum numbers of the $Z_b$ resonances $I^G(J^P)=1^+(1^+)$, this symmetry can be applied to other channels related by HQSS, $I^G(J^P)=1^-(J^+)$ with $J=0$, $1$, and $2$. In this case there should exist four new isovector resonances~\cite{mv11} at the thresholds of $B \bar B$, $B^* \bar B$ and $B^* \bar B^*$ that are inaccessible for observation in the current experimental setting. 

In summary. The weak coupling of the resonance $Z_b(10650)$ to the channel $B^* \bar B + B \bar B^*$ indicated by the data~\cite{bellebb2} implies that in the $Z_b$ states the forces between the heavy mesons largely do not depend on the total spin of the light quark-antiquark pair. Unlike the symmetry of the heavy quark spin, well substantiated in QCD, this approximate light quark spin symmetry is rather unexpected. In particular the effect of the interaction of the light quark spins is likely significantly smaller than could be estimated from the pion exchange between the heavy mesons. The reasons for this behavior are not clear, and it is not known to what extent it is universal across states of the heavy meson pairs with different quantum numbers. However if confirmed for the $Z_b$ mesons, LQSS in combination with HQSS would imply existence of four new isovector resonances at the heavy meson pair thresholds. 

This work is supported in part by U.S. Department of Energy Grant No.\ DE-SC0011842.

\end{document}